\documentclass[prd,preprint,nofootinbib,showpacs]{revtex4}
\usepackage{graphicx}
\usepackage{bm}

\newcommand{\Z}{\mathbf{Z}_2}
\newcommand{\be}{\begin{equation}}
\newcommand{\ee}{\end{equation}}
\newcommand{\bea}{\begin{eqnarray}}
\newcommand{\eea}{\end{eqnarray}}
\newcommand{\N}{\mathcal{N}}
\renewcommand{\b}[1]{\bar{#1}}

\newcommand{\Del}{\nabla}
\newcommand{\del}{\partial}
\newcommand{\bi}{\bar{\imath}}

\renewcommand{\ap}{\alpha^\prime}
\newcommand{\ket}[1]{| #1 \rangle}
\newcommand{\tr}{\mathnormal{tr}}
\newcommand{\ot}{\overline{\Theta}}
\newcommand{\da}{\dot{\alpha}}
\newcommand{\re}{\mathnormal{Re}}

\begin{document}
\preprint{hep-th/0208032}

\title{BPS States of Strings in 3-Form Flux}

\author{Mariana Gra\~na}
\email{mariana@vulcan.physics.ucsb.edu}
\affiliation{Department of Physics\\ University of California\\ 
Santa Barbara, CA 93106}

\author{Andrew R. Frey}
\email{frey@vulcan.physics.ucsb.edu}
\affiliation{Department of Physics\\ University of California\\ 
Santa Barbara, CA 93106}

 
\begin{abstract}
We count the BPS states of strings in uniform 3-form fluxes, using 
supersymmetric quantum mechanics derived from the $\kappa$-symmetric
action for D-branes.  This problem is relevant to the stringy physics of
warped compactifications. We work on a type IIB $T^6/\mathbf{Z}_2$ 
orientifold with imaginary self-dual, quantized,  3-form flux. 
Ignoring the orientifold projection,
the number of short multiplets living on a single string is the square 
of the units of 3-form flux 
present on the torus; the orientifold
removes roughly half of the multiplets.  We review the 
well-known case
of a superparticle on $T^2$ as a pedagogical example.
\end{abstract}

\date{\today}

\maketitle

\section{Introduction}\label{s:intro}

In recent years, following the realization of 
\cite{Randall:1999ee,Randall:1999vf} that ``warped
product'' compactifications can lead to a solution of the hierarchy problem,
string and M theory compactifications with nonvanishing fluxes on the internal 
manifold \cite{Becker:1996gj} 
have garnered much interest
\cite{Mayr:2000zd,Gukov:1999ya,Dasgupta:1999ss,Greene:2000gh} 
(see 
therein for related references).  Of particular interest is the type IIB 
version of \cite{Becker:1996gj} (discussed from the point of view of 
supersymmetry in  
\cite{Grana:2000jj,Gubser:2000vg}), which has been applied to Calabi-Yau 
compactifications with orientifolds \cite{Giddings:2001yu,Becker:2002nn}.
In these compactifications, the 3-form $G=F-\tau H$ (with $\tau = C+i/g_s$) 
must be imaginary self-dual on the internal manifold, $\star_6 G=iG$.

The simplest examples of the warped Calabi-Yau compactifications are the
$T^6/\Z$ form studied in \cite{Verlinde:1999fy,Kachru:2002he,Frey:2002hf},
where, with different 3-form backgrounds, it is possible to have any degree
of supersymmetry up to $\N=4$ ($\N=4$ corresponds to no 3-form flux).  For smaller degrees of SUSY, the low energy
physics is described by the superHiggs effect in supergravity 
\cite{Frey:2002hf,Andrianopoli:2002rm,Andrianopoli:2002mf,
Andrianopoli:2002aq,D'Auria:2002tc,Ferrara:2002bt},
so the SUSY breaking is necessarily accompanied by gauge symmetry breaking.

Frey and Polchinski \cite{Frey:2002hf}
studied the U-dualities of the $\N=3$ string theory compactifications and
found a dilemma related to the superHiggs effect, as follows. 
D5- and NS5-branes wrapped on the torus carry magnetic charges for the
broken vectors and are confined;
the minimal unbroken magnetic charges are therefore bound states of the two 
types of 5-branes.  On the other hand, the minimal electric charges are 
individual D- and F-strings, so there appear to be twice as many electric
BPS multiplets as magnetic for each unbroken symmetry.  In such a case, 
it would appear difficult to have the usual electric-magnetic duality 
transformation for an effective 4D gauge theory.  One suggested 
resolution is that the combined number of BPS states of the two strings
is equal to the number of BPS states of the 5-brane bound state.

In this paper, we carry out the first steps needed to compare the number
of electric and magnetic BPS states by carrying out the ground state
quantization of a D-string in 3-form flux.  To that end, we find the action for the D-string in background fluxes from the known 
supersymmetric and $\kappa$-symmetric Born-Infeld and Wess Zumino actions \cite{Cederwall:1997ri,Bergshoeff:1997tu}, expanded to second order in the bosonic and fermionic world-volume coordinates. For simplicity, 
we will focus on the $\N=3$
background discussed in \cite{Frey:2002hf}, but the techniques we
use are readily generalizable to the other  models described in 
\cite{Giddings:2001yu,Kachru:2002he}.  Also, because of the RR flux and
the finite fixed value of the string coupling, little is known about
such compactifications beyond the low energy supergravity.  This 
calculation sheds a little light on the stringy additions to the spectrum.

The paper is organized as follows.
In section \ref{s:particle}, we review the
simple example of a superparticle on a $T^2/\Z$ orbifold, and in 
section \ref{s:dqm} we describe our supergravity background and
find the action and Hamiltonian of a D-string in that background.  We 
then find the supercharges of the quantum mechanics on the string in 
section \ref{s:susy}, followed by counting the BPS ground states
in section \ref{s:wavefunction}.  Finally, we comment on states isolated at
singularites, bound states, and the number of BPS states of an F-string 
in section
\ref{s:extension}, and section \ref{s:conclude} gives a concluding 
discussion.  Some conventions are given in appendix \ref{a:conventions}, and the size of the corrections to our model are 
shown in appendix \ref{a:assumptions}.

\section{Example of particle on $T^2$}\label{s:particle}

The counting of the spectrum of BPS states for the D-string on the
$T^6/{\bf Z}_2$ orientifold with constant fluxes will turn out to be
very similar to the case of spin 1/2 particle on a torus subject to a
constant perpendicular magnetic field, so in this section we review
the latter, simpler, case (\textit{cf.} 
\cite{Fradkin:1991nr} and references therein).  
We work in the context of supersymmetric
quantum mechanics and calculate a Witten index \cite{Witten:1982df}.

The supersymmetric Lagrangian for a spin 1/2 particle with mass $m$
and charge $e$ is 
\be {\mathcal L}= \frac{m}{2}\dot{\vec{x}}^2 + e
\vec{A} \cdot \dot{\vec{x}} + \frac{i}{2} \psi_1 \dot{\psi_1}+
\frac{i}{2} \psi_2 \dot{\psi_2} - i\frac{e}{m} \psi_1 \psi_2 B
\label{lagr}
\ee where $\psi_1$ and $\psi_2$ are two real Grassman variables,
obeying \be\label{anticom} \{\psi_i,\psi_j\}= \delta_{ij}\ . \ee
We should note here that we are using conventions such that 
$(\psi\phi)^\dagger=\phi^\dagger\psi^\dagger$, leading to the real
anticommutator (\ref{anticom}).\footnote{The anticommutator follows from
the theory of systems with constraints, as detailed in
\cite{Casalbuoni:1976tz}; this is important for the D-string, since it
gives constants in the anticommutator.  Note especially that the Dirac 
quantization does not give a factor of $1/2$ in the anticommutator.}
The Lagrangian (\ref{lagr}) is invariant under the 
supersymmetry transformation \be \delta x^i=i \psi_i \epsilon \qquad
\delta \psi_i= -m \dot{x}^i \epsilon
\label{susytransf}
\ee
up to a total derivative, as can be verified by a simple calculation.

We find it convenient to work in the operator formalism, with
Hamiltonian \be\label{ham} H=
\frac{1}{2m}\left(\vec{p}-e\vec{A}\right)^2 +  i\frac{e}{m} \psi_1
\psi_2 B\ . \ee
Then the  supersymmetry (\ref{susytransf}) is generated by the supercharge
operator \be Q=i\left(\vec{p}-e\vec{A}\right)\cdot
\psi
\label{supercharge}
\ee satisfying $Q^2=mH$.

In terms of the complex combinations 
\be z=\frac{1}{2} (x^1 + i x^2)\ ,\qquad
\psi_{\pm}=\frac{1}{2}(\psi_1 \pm i \psi_2)\ , \label{comp}\ee 
the supercharge
(\ref{supercharge}) can be rewritten as \be Q= i(p-eA)_z \psi_+ +i 
(p-eA)_{\bar{z}} \psi_-
\ .\label{superz}
\ee
Supersymmetric ground states should be annihilated by the
supercharge. If we start with a state $\ket{-}$ that is annihilated by
$\psi_-$, then, from (\ref{superz}), supersymmetric wave functions 
obey 
\be (p - eA)_z \phi_- = 0
\ .\label{eqphi}
\ee

 Let's use a gauge for the potential in which 
\be\label{EMpotential} A_1=-\frac{1}{2}B
x^2\ , \quad  A_2=\frac{1}{2}B x^1\ \Rightarrow\quad A_z=-iB\b{z}\ . \ee 
This is periodic up to a gauge transformation, 
\be \label{gaugetr}
A_i(x^1+2\pi
R, x^2)= A_i(x^1,x^2)+\partial_i \lambda_1, \qquad  A_i(x^1, x^2+2\pi
R)= A_i(x^1,x^2)+\partial_i \lambda_2 \ee 
with $\lambda_1=B \pi R\,
x^2$ and $\lambda_2=-B \pi R\, x^1$. When going around the torus, the
wave function picks up a phase determined by these gauge
transformations, \be \phi(x^1+2\pi R, x^2)=e^{ie\lambda_1}
\phi(x^1,x^2), \qquad  \phi(x^1, x^2+2\pi R)=e^{ie\lambda_2}
\phi(x^1,x^2)\ .
\label{period}
\ee
Single-valuedness of the wavefunction implies that 
the magnetic field is quantized,  \be B= \frac{n}{2\pi e R^2}
\label{magnetic}
\ee where $n\in {\bf Z}$, and $R$ is the radius of the square torus.
If there is a $\Z$ orbifold, any flux of the form (\ref{magnetic}) is still allowed, but some
of the fixed points will carry flux, also.  These special fixed points 
will not affect our discussion below, since the translations and reflections
automatically give the wavefunction the correct boundary conditions at
those fixed points.
This is discussed in detail in \cite{Frey:2002hf}.

>From (\ref{eqphi}), the wave function should satisfy \be
\frac{\partial \phi_-}{\partial z}= eB \bar{z}\phi_- \ee whose
solution is \be \phi_-=e^{eB |z|^2} F(\bar{z})
\ .\label{sol1}
\ee The periodicity conditions (\ref{period}), written in complex
coordinates, are \be \phi(z+\pi R, \bar{z}+ \pi R)=e^{e B \pi
(z-\bar{z})}\phi(z,\bar{z}), \qquad \phi(\bar{z}+i\pi R, \bar{z}-i \pi
R)=e^{-i e B \pi (z+\bar{z})}\phi(z,\bar{z})
\ .\label{periodzcond}
\ee Inserting (\ref{sol1}), we get from the first condition in
(\ref{periodzcond}) \be F(\bar{z}+ \pi R)= e^{-eB2 \pi R 
\bar{z}-eB\pi^2R^2}F(\bar{z})=e^{-\frac{n}{R}\bar{z}-\frac{n\pi}{2}}
F(\bar{z})
\ ,\label{periodF}
\ee where in the last equality we used the quantization condition
(\ref{magnetic}). Defining \be \label{defG} F(\bar{z})= e^{-\frac{n}{\pi
R^2}\bar{z}^2}G(\bar{z}) \ee the condition (\ref{periodF}) implies
that $G(\bar{z})$ is periodic, with period $\pi R$. Writing
$G(\bar{z})$ as a sum of Fourier modes, we get for the wave function
\be \phi_-= e^{\frac{n}{2\pi R^2} |z|^2-\frac{n}{2\pi R^2}\bar{z}^2}
\sum_{-\infty}^{\infty} C_m e^{\frac{2i}{R}m \bar{z}}\ .
\label{finsol}
\ee The second periodicity condition in (\ref{periodzcond}) implies
the recursion relation \be\label{recurs-} C_{m+n}=C_m e^{\pi (n+2m)}\ . \ee 
So only $n$ of the coefficients $C_m$ are
free. With this recursion relation, the series in (\ref{finsol}) converges if $n<0$. So, for a
magnetic field in the negative $3$ direction, there are $n$ ground
states.

If instead we had started with a ground state annihilated by $\psi_+$,
then the wave function would have been \be \phi_+= e^{-\frac{n}{2\pi
R^2} |z|^2+\frac{n}{2\pi R^2}z^2} \sum_{-\infty}^{\infty} C_m
e^{\frac{2i}{R}mz}
\label{finsol+}
\ee and the constraint on the components $C_m$ turns out to be \be
C_{m+n}=C_m e^{-\pi (n+2m)}\ , \label{recurs+}\ee 
so in this case the sum in (\ref{finsol+})
converges for positive $n$.

Let's take a minute to note the relation of the wavefunctions to the
well-known theta functions on the torus (see chapter 7 of 
\cite{Polchinski:1998rq} for a review).  For example, with $n>0$, the
recursion relation (\ref{recurs+}) has solution $C_m = D e^{-\pi m^2/n}$
with constant $D$,
so the wavefunction (\ref{finsol+}) is a Gaussian times the 
theta function
\be\label{thetafunct}
\phi_+=  \sum_{k=0}^{n-1} D_k \exp\left[-\frac{n}{2\pi
R^2} |z|^2+\frac{n}{2\pi R^2}z^2\right]
\vartheta\left[\begin{array}{c}k/n\\0\end{array}\right] 
\left(\frac{nz}{\pi R},
in\right) \ .\ee
(We use the notation of \cite{Polchinski:1998rq}.)  The sum now has $n$
independent coefficients $D_k$.

A spin $1/2$ particle in a constant perpendicular magnetic field on a
torus has then a finite number of ground states, given by the number
of units of magnetic flux.  On a $\Z$ orbifold, we must have $z\simeq -z$,
or $C_m=C_{-m}$, which force relationships between the $D_k$ coefficients.  
Thus the number of supersymmetric ground states becomes 
$(n+1)/2$ for $n$ odd and $n/2 +1$ for $n$ even.\footnote{It seems naively
that the $T^2/\Z$ orbifold can be smoothly deformed to $S^2$, but in that
case an even number of flux units would give $n/2$ states.  Apparently
the orbifold limit of the sphere is singular; we leave the resolution of
this problem as an exercise for the reader.}
If we think of the states
$\ket{\pm}$ as respectively bosonic and fermionic, the number of ground 
states is therefore the Witten index $\tr (-1)^F$.

\section{D-string quantum mechanics}\label{s:dqm}
In this section, we find the Lagrangian and Hamiltonian for a D-string
extended along one direction of a $T^6/\Z$ orientifold in imaginary 
self-dual 3-form flux, including the fermionic degrees of freedom.  

\subsection{Supergravity background}\label{ss:back}
We start by describing the supergravity background, which was given in 
\cite{Kachru:2002he,Frey:2002hf}.  In this section, we use Greek
indices for spacetime coordinates and Latin for torus coordinates.
For a general amount of SUSY up to
$\N = 4$ ($\N=4$ corresponding to $G_3=0$), the fields have the form
\bea
ds^2&=&Z^{-1/2}\eta_{\mu\nu}dx^\mu dx^\nu +Z^{1/2}d\hat s_6^2\ ,
\label{metric10}\\
\tau&=& C+\frac{i}{g_s}=\mathnormal{constant}\label{dilaton}\\
\tilde F_5&=& (1+\star )d\chi_4\ ,\quad \chi_4 = \frac{1}{Z g_s} dx^0 \wedge
dx^1 \wedge dx^2 \wedge dx^3\label{form5}\\
\star_6 G_3&=&iG_3\\
-\hat\Del^2 Z&=&(2\pi)^4\ap{}^2g_s \hat\rho_3+\frac{g_s}{2}|\hat G|^2
\label{warp}\ .
\eea
Here, $d\hat s^2=g_{m\b n}dy^md\b y^{\b n}$ ($m,n=4,\ldots 9$) 
is the unwarped metric on the torus, which we describe 
with the complex pairs $y^m=(x^m+t^{mn}x^n)/2$, with here
$m=4,5,6$, $n=7,8,9$.  The
K\"ahler moduli $g_{m\b n}$ are free, so we choose them to give the unit
metric for simplicity.  
The complex structure is given by the period matrix $t$ \cite{Kachru:2002he}, 
which will
be fixed by the 3-forms; we will consider a case with $t^{mn}=i\delta^{m+3,n}$.
We will find it convenient to have the proper radii implicit in the
coordinates: $x^{4,7}\simeq x^{4,7}+2\pi R_1$, $x^{5,8}\simeq x^{5,8}
+2\pi R_2$, $x^{6,9}\simeq x^{6,9}+2\pi R_3$.
Additionally, the dilaton-axion is fixed by the fluxes, and we will study cases
with vanishing RR scalar.
We should note also
that (\ref{warp}) implies that $Z=1+\mathcal{O}(\ap{}^2/R^{4})$ in the large
radius limit.

As in \cite{Frey:2002hf}, these values for the fixed moduli are consistent
with constant 3-forms
\be\label{quant3form}
H_{mnp}= \frac{\ap}{2\pi R_1R_2R_3} h_{mnp}\ ,\quad 
F_{mnp}= \frac{\ap}{2\pi R_1R_2R_3} f_{mnp}\ ,\quad h_{mnp},f_{mnp}\in
\mathbf{Z}
\ee
given by 
\bea
h_1&=&h_{456}=-h_{489}=-h_{759}=-h_{786}\ ,\nonumber\\
h_2&=&h_{789}=-h_{756}=-h_{486}=-h_{459}\ ,\nonumber\\
f_1&=&f_{456}=-f_{489}=-f_{759}=-f_{786}\ ,\nonumber\\
f_2&=&f_{789}=-f_{756}=-f_{486}=-f_{459}\label{fluxes}\ .
\eea
As in the case of the superparticle, these integers can be odd, as long
as some of the orientifold fixed planes carry flux.  These will not affect
the quantization of the string \cite{Frey:2002hf}.
With vanishing RR scalar, imaginary self-duality gives
\be\label{selfduality}
f_2=\frac{h_1}{g_s}\ ,\quad f_1=-\frac{h_2}{g_s}\ .\ee
This background has 4D $\N=3$ supersymmetry.  In this background, a D-string
and orthogonal F-string carry the same central charge (for example, a D-string
wrapped on $x^4$ and an F-string on $x^7$) and fall in short multiplets
of the $\N=3$ theory.  These preserve 4 of the 12 supercharges preserved 
by the background and have 16 states;
we will be counting the number of multiplets associated with each string.
We should caution the reader that it is impossible to be certain that
these BPS states are truly D- or F-strings, since we are working at
$g_s\sim 1$, but we see in this paper that the physics of D-strings appears
to make sense even at this strong coupling.

\subsection{Action and Hamiltonian}\label{ss:action}
Here we find the supersymmetric action for a D-string wrapped (without loss
of generality) on $x^4$ in the self-dual 3-form flux described in section
\ref{ss:back}.  Because the number of BPS states is stable under small
perturbations (in the 4D $\N =3$ case we consider, for example, a long
multiplet is 4 times the size of the BPS multiplet), we will work in the 
large radius limit of the compactification and ignore the warp factor.  
Also, because motion along the string is quantized by the compactification,
we will set those derivatives to zero.  Finally, our D-string will not 
intersect any D3-branes, so we ignore states associated with 1-3 strings.  
We will estimate the corrections to
our simple model 
in appendix \ref{a:assumptions}.  Because we are studying only one type
of electric charge in the 4D theory, the D-string will not be wound in
any other direction or carry any dissolved F-strings.

The supersymmetric action for a D1-brane in background fluxes was
worked out  in \cite{Cederwall:1997ri,Bergshoeff:1997tu}. It looks like the
bosonic Dirac-Born-Infeld and Wess-Zumino actions, but the spacetime
fields live in superspace.
\begin{equation}
S=-\frac{1}{2\pi\ap}\int d^2\zeta e^{- \bm{\Phi}}\sqrt{-\det \left(
\bm{g}_{\mu\nu} 
+ \bm{\mathcal{F}}_{\mu\nu}\right)}+\frac{1}{2\pi\ap}\int 
e^{\bm{\mathcal{F}}} \wedge\bm{C}\ .
\label{susyaction}
\end{equation}
The fields in boldface are superfields; as usual,
\begin{equation}
\bm{\mathcal{F}}=2\pi\ap F-\bm{B}\ ,\ \mathnormal{and}\ 
\bm{C}=\oplus_n \bm{C}_{(n)}
\label{somedefs}\end{equation}
is the collection of all RR potentials pulled back to the
world-volume.  The expansions of the superfields in terms of
components fields was developed in \cite{deWit:1998tk}, \cite{Millar:2000ib}, 
and \cite{Grana:2002tu} for 11-dimensional, IIA, and IIB  supergravities
respectively, using a method known as gauge completion. The expansions
of the fields that we will need, as well as our conventions, are listed
in appendix \ref{a:conventions}. 

Without getting into algebraic details, the action (\ref{susyaction}) 
for a D-string in our background is
\bea
S&=&-\frac{1}{2\pi \ap g_s}\int d^2\zeta \left[ \left(1-\mathcal{F}_{04}^2
\right)^{1/2}-\frac{1}{2}  \left(1-\mathcal{F}_{04}^2\right)^{-1/2}
(\dot{x}^m)^2-g_s C_{m4}\dot{x}^m\right.\nonumber\\
&&\left.+i\frac{g_s^{1/2}}{2} \ot \Gamma^0 \dot{\Theta} 
+i\frac{g_s^{3/2}}{16} \ot \Gamma_0{}^{mn} \Theta F_{4mn} 
+i\frac{g_s^{1/2}}{48} \ot \Gamma^{mnp} \Theta H_{mnp}-i\frac{g_s^{1/2}}{16}  
\ot \Gamma^{4mn} \Theta H_{4mn}\right]
\label{action1}
\eea
where we work to second order in the world-volume coordinates and 
fermions.\footnote{$F_{4mn}$ in the second fermionic term would be 
$F'_{4mn} = F_{4mn}-CH_{4mn}$, and also the Chern-Simons term would couple
the velocity $\dot{x}^m$ to $C_{m4}+C\mathcal{F}_{m4}$, if the RR scalar is nonvanishing.}
As noted above, there are corrections to this action, arising from the 
expansion of the D-brane action; we consider these
in appendix \ref{a:assumptions}.

As discussed in appendix \ref{a:conventions}, the fermion $\Theta$ is a
10D (Majorana-Weyl) superspace coordinate; 
let us now expand it in terms of 2D spinors.
We do so by noting that
the 10-dimensional gamma matrices can be decomposed into  
$SO(1,1)\otimes SO(8)$ pieces as
\be\label{gammas}
\Gamma^{\parallel}=\gamma^{\parallel}\otimes 1 , \qquad \Gamma^{\perp}= 
\gamma_{(2)}\otimes \gamma^{\perp}\ ,
\ee
where ``$\parallel$'' and ``$\perp$'' mean along the D-string and
perpendicular to it; $\Gamma$ is a 32x32 Dirac matrix,
$\gamma^{\parallel}$ and $\gamma^{\perp}$ are its 2x2 and 16x16
blocks, and $\gamma_{(2)}$ is the chirality matrix in $SO(1,1)$.
Therefore, a  Majorana-Weyl spinor $\Theta$ can be decomposed into 
\be
\Theta= \ket- \otimes \psi^{\alpha} u_{\alpha} \oplus \ket+ \otimes 
\phi^{\dot{\alpha}} v_{\da}
\ ,\label{decomp}
\ee
where $\ket{+}$ and $\ket{-}$ are the eigenfunctions of $\gamma_{(2)}$, 
and $\alpha$ and 
$\da$ are indices in the ${\bf 8}$ and ${\bf 8'}$ representations of $SO(8)$.
The $\psi$s and $\phi$s are 2D Majorana-Weyl fermions. 
The spin raising (lowering) gamma matrices are the (anti)holomorphic
gamma matrices defined with respect to the coordinates
\bea
z^{0\pm}=\frac{1}{2} (\pm x^0+x^4)&,& z^{1}=\frac{1}{2} 
(x^1+ ix^2)\ , \quad z^{2}=\frac{1}{2} (x^3+ ix^7)\ ,\nonumber\\
\quad z^{3}=\frac{1}{2} (x^5+ ix^8)&,&z^{4}=
\frac{1}{2} (x^6+ ix^9)\ .\label{zcomplex}
\eea
We will label the complex coordinates 1,2,3,4 with indices $i,j,\ldots$.

Using the basis (\ref{so8basis}) for the Majorana-Weyl spinors and integrating along the string, 
the fermionic Lagrangian from eq (\ref{action1}) can be written
\bea
L_f &=& - i\frac{R_1}{2\ap g_s^{1/2}}\left[
\psi^{\alpha}\dot{\psi}^{\alpha} +  \phi^{\da}\dot{\phi}^{\da}
\right.\nonumber\\
&&\left. + \frac{\ap}{2\pi R_1R_2R_3}h_1
\left(\psi^1\psi^4+\psi^2\psi^3+\phi^4\phi^1+\phi^3\phi^2+\psi^1\phi^3+
\phi^4\psi^2+\phi^1 \psi^3+\psi^4\phi^2\right)\right. \nonumber\\
&& \left. +\frac{\ap}{2\pi R_1R_2R_3}
h_2\left(\psi^3\psi^1+\psi^2\psi^4+\phi^1\phi^3+\phi^4\phi^2
+\psi^1\phi4+\psi^2\phi^3+\phi^2\psi^3+\phi^1\psi^4\right)\right]\ .
\label{fermions}\eea
As well as some algebraic simplification, arriving at 
this result requires 3-form self-duality to relate the NSNS and RR fluxes as in (\ref{selfduality}).
We should note that the fermions $\psi^\alpha$ and $\phi^{\da}$ with 
$\alpha,\da =5,6,7,8$ enter only through their kinetic terms.

Now to convert to the Hamiltonian formalism, we start by finding the canonical
momentum for the world-volume gauge field $F=dA$.  Following the discussion
in \cite{Witten:1982df}, the Wilson lines are periodic variables, so the
momentum 
\be\label{gaugemomentum}
p_A = \frac{2\pi R_1}{g_s}
\frac{\mathcal{F}_{04}}{[1-\mathcal{F}_{04}^2]^{1/2}}\ee
is quantized in units of $2\pi R_1$.  Further, it is this canonical 
momentum (up to constants) which couples the D-string to $B$, so $p_A$
and therefore $\mathcal{F}$,
\emph{not} the gauge field strength $F$, vanish for a D-string with no
F-string charge.  This issue is somewhat more complicated in the presence
of the RR scalar $C$, but we have now removed the NSNS flux from the 
problem.

The canonical momenta for the collective coordinates now simplify to
\be\label{momenta}
p_m = \frac{R_1}{\ap g_s}\dot x_m +\frac{R_1}{\ap} C_{m4}\ , \ee
as in the usual quantum mechanics with a gauge field proportional to $C_{m4}$.  Thus, the total Hamiltonian is a
constant mass $m=R_1/\ap g_s$ and a dynamical Hamiltonian
\bea
H&=& \frac{1}{2m}(\vec{p}-\vec{A})^2 + i\frac{C}{2}h_1
\left(\psi^1\psi^4+\psi^2\psi^3+\phi^4\phi^1+\phi^3\phi^2+\psi^1\phi^3+
\phi^4\psi^2+\phi^1 \psi^3+\psi^4\phi^2\right) \nonumber\\
&&+ i\frac{C}{2}h_2
\left(\psi^3\psi^1+\psi^2\psi^4+\phi^1\phi^3+\phi^4\phi^2
+\psi^1\phi4+\psi^2\phi^3+\phi^2\psi^3+\phi^1\psi^4\right)\label{hamil}
\eea
with $C=(2\pi R_2R_3g_s^{1/2})^{-1}$.  The gauge field is defined as
$A_m = (R_1/\ap)C_{m4}$.

\section{Supercharges}\label{s:susy}

In this section, we demonstrate that the Hamiltonian (\ref{hamil}) is, in
fact, supersymmetric, and we identify the 4 supercharges that belong
to the unbroken supersymmetries of the $\N =3$ 4D effective theory.  We 
will proceed by first finding the spacetime supersymmetries that leave both
the background (see section \ref{ss:back}) and the BPS states of the string
invariant by starting with the 10D theory.  
We will then relate those to world-volume supercharges,
which should be of the form given in eq. (\ref{supercharge}), 
$Q\sim i(p-A)\psi$.  Finally, we will check that these do actually commute 
with the Hamiltonian.

In the IIB string theory, the supercharges are Majorana-Weyl spinors with
positive chirality, one coming from each side of the string, and the
superalgebra contains NSNS and RR charges as tensorial central charges
(see \cite{Polchinski:1998rr}).  In this notation, the supercharges 
preserved by a BPS state are given by $\b \varepsilon_1 Q + \b \varepsilon_2
\tilde Q$ ($\varepsilon_{1,2}$ have negative chirality), where 
\be\label{superalgebra}
\left\{ \left[\begin{array}{c} Q\\ \tilde Q\end{array}\right],
\left[\begin{array}{cc} \b Q& \b{\tilde Q}\end{array}\right]\right\}
\left[\begin{array}{c}\varepsilon_1\\ \varepsilon_2\end{array}\right]=0
\ .\ee
In contrast, the supercharges of supergravity backgrounds (as in
\cite{Grana:2000jj,Gubser:2000vg}) are denoted following the conventions of
\cite{Schwarz:1983wa,Schwarz:1983qr}.  In this form, the supersymmetry
parameters are given by a single negative chirality Weyl spinor, so that the
preserved supercharges are given by $\b\varepsilon \mathcal{Q}+
\b\varepsilon^*\mathcal{Q}^*$.  To relate these two formalisms, we consider
the supersymmetries preserved by a D3-brane in flat spacetime.  In the form 
of (\ref{superalgebra}), it is easy to see that $\varepsilon_2=i\gamma_{(4)}
\varepsilon_1$, whereas the supergravity formalism gives 
$\varepsilon=\gamma_{(4)}\varepsilon$ (see, for example, 
\cite{Grana:2000jj,Gubser:2000vg}), where $\gamma_{(4)}$ is the chirality
along the D3-brane.  The conventions agree if we take
$\varepsilon = \varepsilon_1 - i\varepsilon_2$ and $\mathcal{Q}=(Q+
i\tilde Q)/2$.

The supersymmetries preserved by our background (section \ref{ss:back}) 
are expressed conveniently by $SO(3,1)\otimes SO(6)$ decomposition 
$\varepsilon^A = \eta\otimes \chi^A$, where $\chi^A$ are 3 of the 4 negative
chirality spinors in 6D (those that don't have the three spins parallel)  \cite{Frey:2002hf}.  
We can re-write these in the $SO(1,1)\otimes SO(8)$ basis
(\ref{so8basis}) as
\bea
\varepsilon^1&=& \epsilon^1_1 \left[ \ket{-}\otimes (u_1-iu_2)
-\ket{+}\otimes (v_1-iv_2)\right] +\epsilon^1_3 \left[ \ket{-}\otimes
(u_3-iu_4)+\ket{+}\otimes (v_3-iv_4)\right]\ ,\nonumber\\
\varepsilon^2&=& \epsilon^2_5 \left[ \ket{-}\otimes (u_5-iu_6)
-\ket{+}\otimes (v_5-iv_6)\right] +\epsilon^2_7 \left[ \ket{-}\otimes
(u_7+iu_8)+\ket{+}\otimes (v_7+iv_8)\right]\ ,\nonumber\\
\varepsilon^3&=& \epsilon^3_5 \left[ \ket{-}\otimes (u_5+iu_6)
+\ket{+}\otimes (v_5+iv_6)\right] +\epsilon^3_7 \left[ \ket{-}\otimes
(u_7-iu_8)-\ket{+}\otimes (v_7-iv_8)\right]\qquad\label{n3susy}
\eea
with complex Grassman numbers $\epsilon^A_\alpha$.  

Now, let's intersect these with the supersymmetries preserved by the D-string. Using the superalgebra with central charges, it is easy to find that a
D-string wrapped on $x^4$ has supersymmetries $\varepsilon_1=\gamma_{(2)}
\varepsilon_2$ \cite{Polchinski:1998rr}.  Then, using $\varepsilon=\varepsilon_1-i\varepsilon_2$,  we have 
$\gamma_{(2)}\varepsilon=-i\varepsilon^*$, so the coefficients of the $u$
spinors must satisfy $\epsilon=i\epsilon^*$ and the coefficients of $v$
spinors satisfy $\epsilon=-i\epsilon^*$.  We can obtain spinors that 
satisfy these constraints by taking linear combinations of the latter two
spinors (\ref{n3susy}) such that $\epsilon^2_5=\pm\epsilon^3_5$ and
$\epsilon^2_7=\pm\epsilon^3_7$.  In the end, we find 4 different one 
component spinors $\varepsilon^A$ with
\bea
\varepsilon^1=\epsilon^1(\ket{-}\otimes u_5+i\ket{+}\otimes v_6)&,&
\varepsilon^2=\epsilon^2(\ket{-}\otimes u_6-i\ket{+}\otimes v_5)\ ,
\nonumber\\
\varepsilon^3=\epsilon^1(\ket{-}\otimes u_7+i\ket{+}\otimes v_8)&,&
\varepsilon^4=\epsilon^4(\ket{-}\otimes u_8-i\ket{+}\otimes v_7)\ .
\label{n3-d4susy}\eea
The coefficients are $\epsilon^A = \epsilon^{\prime A} e^{i\pi/4}$ with
$\epsilon^{\prime A}$ real.

Now we can actually find the worldvolume supercharges.  As discussed in
\cite{Bergshoeff:1997kr,Kallosh:1998ky}, the worldvolume 
supersymmetries are not given simply by their 
action on the spacetime fields (including $\Theta$ as a superspace coordinate)
because that transformation would in general change the $\kappa$-symmetry
gauge.  The supersymmetry transformations of the worldvolume fields,
including a $\kappa$ transformation to keep the same gauge, were found
in \cite{Kallosh:1998ky}.  However, we will not follow this approach.  
Instead, we will take the ansatz
\be\label{Qansatz}
-\frac{1}{\sqrt{2}} Q^A\epsilon^A = i(p-A)_i \b\Theta \Gamma^i 
\re\left( \varepsilon^A\right) 
+i(p-A)_{\bi} \b\Theta \Gamma^{\bi}\re\left(\varepsilon^A\right)
\ee
for the supercharges.  We know that the real part of the spinor should be
used because the supercharges should be real (as the spacetime ones are),
and we follow \cite{Bergshoeff:1997kr} in using the spacetime supersymmetry
parameters.  The prefactors are included for convenience.

Using this ansatz, we can write the supercharges in terms of the 2D fermions
$\psi,\phi$ and complex coordinates $z$ as
\bea
Q^1&=& ip_1 (-i\psi_7-\psi_8+\phi_7-i\phi_8)+i p_{\b 1} (i\psi_7-\psi_8+\phi_7+
i\phi_8)\nonumber\\
&&+ip_2(-i\psi_5+\psi_6-\phi_5+i\phi_6)
+ip_{\b 2}(i\psi_5+\psi_6-\phi_5-i\phi_6)\nonumber\\
&&+i(p-A)_3(-i\psi_3-\psi_4-\phi_3
+i\phi_4)+i(p-A)_{\b 3}(i\psi_3-\psi_4-\phi_3-i\phi_4)\nonumber\\
&&+i(p-A)_4(i\psi_1+\psi_2+\phi_1
-i\phi_2)+i(p-A)_{\b 4}(-i\psi_1+\psi_2+\phi_1+i\phi_2)\ ,\nonumber\\
Q^2&=& ip_1 (-\psi_7+i\psi_8-i\phi_7-\phi_8)+ip_{\b 1} (-\psi_7-i\psi_8+i\phi_7
-\phi_8)\nonumber\\
&&+ip_2(-\psi_5-i\psi_6-i\phi_5-\phi_6)+ip_{\b 2}
(-\psi_5+i\psi_6+i\phi_5-\phi_6)\nonumber\\
&&+i(p-A)_3(-\psi_3+i\psi_4+i\phi_3
+\phi_4)+i(p-A)_{\b 3}(-\psi_3-i\psi_4-i\phi_3+\phi_4)\nonumber\\
&&+i(p-A)_4(-\psi_1+i\psi_2+i\phi_1
+\phi_2)+i(p-A)_{\b 4}(-\psi_1-i\psi_2-i\phi_1+\phi_2)\ ,\nonumber\\
Q^3&=& ip_1 (i\psi_5+\psi_6-\phi_5+i\phi_6)+ip_{\b 1} (-i\psi_5+\psi_6-\phi_5-
i\phi_6)\nonumber\\
&&+ip_2(-i\psi_7+\psi_8-\phi_7+i\phi_8)
+ip_{\b 2}(i\psi_5+\psi_8-\phi_7-i\phi_8)\nonumber\\
&&+i(p-A)_3(-i\psi_1-\psi_2-\phi_1
+i\phi_2)+i(p-A)_{\b 3}(i\psi_1-\psi_2-\phi_1-i\phi_2)\nonumber\\
&&+i(p-A)_4(-i\psi_3-\psi_4-\phi_3
+i\phi_4)+i(p-A)_{\b 4}(i\psi_3-\psi_4-\phi_3-i\phi_4)\ ,\nonumber\\
Q^4&=& ip_1 (\psi_5-i\psi_6+i\phi_5+\phi_6)+ip_{\b 1} (\psi_5+i\psi_6-i\phi_5
+\phi_6)\nonumber\\
&&+ip_2(-\psi_7-i\psi_8-i\phi_4-\phi_8)+ip_{\b 2}
(-\psi_7+i\psi_8+i\phi_7-\phi_8)\nonumber\\
&&+i(p-A)_3(\psi_1-i\psi_2-i\phi_1
-\phi_2)+i(p-A)_{\b 3}(\psi_1+i\psi_2+i\phi_1-\phi_2)\nonumber\\
&&+i(p-A)_4(-\psi_3+i\psi_4+i\phi_3
+\phi_4)+i(p-A)_{\b 4}(-\psi_3-i\psi_4-i\phi_3+\phi_4)\ .\label{wvsusy}
\eea
It is a straightforward but tedious calculation to show that each of these
commute with the Hamiltonian (\ref{hamil}).  We need to note that canonical
quantization gives the anticommutators $\{\psi_\alpha,\psi_\beta\}=
\{\phi_{\da},\phi_{\dot\beta}\}=\delta_{\alpha\beta}/(mg_s^{1/2})$
\cite{Casalbuoni:1976tz}.
The magnetic field $F=dA$ (where $A$ was defined below Eq.(\ref{hamil})) 
in the $z$ coordinates is also necessary:
\be\label{zmag}
F_{34}=-\frac{1}{\pi R_2R_3 g_s}(h_2-ih_1)\ ,\qquad
F_{\b 3\b 4}=-\frac{1}{\pi R_2R_3 g_s}(h_2+ih_1)\ .\ee
This arises in $[Q,H]$ from commutators $[p,A]$; there are no commutators
mixing holomorphic and antiholomorphic indices.

\section{Supersymmetric ground states}\label{s:wavefunction}

To find the states annihilated by the supercharges (\ref{wvsusy}), 
it's easier to 
work in the complex basis
\be
w^1=\frac{1}{2}(\tilde{x}^5+i \tilde{x}^6) , \qquad w^2=\frac{1}{2}
(\tilde{x}^8+i \tilde{x}^9)
\label{basis}
\ee
where
\bea
\tilde{x}^5=\frac{1}{\sqrt{2h(h-h_2)}}\left(h_1 x^5+(h-h_2)x^8\right)&,& 
\tilde{x}^6=\frac{1}{\sqrt{2h(h-h_2)}}\left(h_1 x^6+(h-h_2)x^9\right)
\nonumber\\
\tilde{x}^8=\frac{1}{\sqrt{2h(h-h_2)}}\left((h-h_2) x^5-h_1x^8\right)&,& 
\tilde{x}^9=\frac{1}{\sqrt{2h(h-h_2)}}\left((h-h_2) x^6-h_1x^9\right)
\label{xtilde}
\eea
and $h=\sqrt{h_1^2+h_2^2}$. In the basis (\ref{basis}), the nonzero 
components of the magnetic field are
\be\label{wmagfield}
F_{w^1\bar{w}^1}=-\frac{1}{\pi R_2 R_3}\frac{ih}{g_s}\, , \qquad 
F_{w^2\bar{w}^2}=\frac{1}{\pi R_2 R_3}\frac{ih}{g_s}\ee
We can use a gauge where the potential is
\bea
A_{w^1}=\frac{1}{2 \pi R_2 R_3} \frac{ih}{g_s} \bar{w}^1 &,& 
A_{\bar{w}^1}=-\frac{1}{2 \pi R_2 R_3}\frac{ih}{g_s} w^1 \nonumber \\
A_{w^2}=-\frac{1}{2 \pi R_2 R_3} \frac{ih}{g_s} \bar{w}^2 &,& 
A_{\bar{w}^2}=\frac{1}{2 \pi R_2 R_3}\frac{ih}{g_s} w^2 \ .
\label{potw}
\eea

The supercharges (\ref{wvsusy}) can be rewritten (up to sign)
\bea
Q^1 &=&(p-A)_{w^1}(\lambda_1-\lambda_4)+ (p-
A)_{\bar{w}^1}(-\bar{\lambda}_1+\bar{\lambda}_4)\nonumber\\
&&+(p-A)_{w^2}(\lambda_2+\lambda_3)+ (p-A)_{\bar{w}^2}
(-\bar{\lambda}_2+\bar{\lambda}_3)+... \nonumber\\
Q^2&=&(p-A)_{w^1}(\bar{\lambda}_2+\bar{\lambda}_3)+ (p
-A)_{\bar{w}^1}(-\lambda_2-\lambda_3)\nonumber\\
&&+(p-A)_{w^2}(-\bar{\lambda}_1+\bar{\lambda}_4)+ (p
-A)_{\bar{w}^2}(\lambda_1-\lambda_4)+... \\
Q^3&=& i(p-A)_{w^1}(\lambda_1+\lambda_4)+ i(p
-A)_{\bar{w}^1}(-\bar{\lambda}_1+\bar{\lambda}_4)\nonumber\\
&&+i(p-A)_{w^2}(-\lambda_2+\lambda_3)+ i(p
-A)_{\bar{w}^2}(-\bar{\lambda}_2-\bar{\lambda}_3)+... \nonumber\\
Q^4&=&i(p-A)_{w^1}(-\bar{\lambda}_2-\bar{\lambda}_3)+ i(p-
A)_{\bar{w}^1}(-\lambda_2-\lambda_3)\nonumber\\
&&+i(p-A)_{w^2}(\bar{\lambda}_1-\bar{\lambda}_4)+ i(p-
A)_{\bar{w}^2}(\lambda_1-\lambda_4)+... 
\label{Qs}
\eea
where $+...$ are terms involving momenta in the noncompact and $x^7$ 
directions, which will give zero when acting on the ground state 
wave-functions. The 
fermions $\lambda_\alpha$ in (\ref{Qs}) are defined in terms of the 
fermions in (\ref{decomp}) as
\bea
\lambda_1&=&\frac{1}{\sqrt{2h(h-h_2)}}\left(h_1 (\psi_2+i\psi_4)+(h-h_2)
(\psi_1+i\psi_3)\right)\nonumber\\
\lambda_2&=&\frac{1}{\sqrt{2h(h-h_2)}}\left((h-h_2) (\psi_2+i\psi_4)-h_1
(\psi_1+i\psi_3)\right)\nonumber\\
\lambda_3&=&\frac{1}{\sqrt{2h(h-h_2)}}\left(h_1 (\phi_2+i\phi_4)+(h-h_2)
(\phi_1+i\phi_3)\right)\nonumber\\
\lambda_4&=&\frac{1}{\sqrt{2h(h-h_2)}}\left((h-h_2) (\phi_2+i\phi_4)-h_1
(\phi_1+i\phi_3)\right)\ .\label{lambdaferms}
\eea
These spinors satisfy the oscillator algebra
\be\label{osc}
\{\lambda_\alpha,\lambda_\beta\}=\{\bar{\lambda}_\alpha, 
\bar{\lambda_\beta}\}=0, \qquad 
\{\lambda_\alpha, \bar{\lambda}_\beta\}=k \delta_{\alpha\beta}
\ee
where $k$ is a real constant that can be absorbed in the definition of the 
spinors. So $\lambda$, $\bar{\lambda}$ are raising and lowering operators.

As in the $T^2$ example, to build the wave functions, we start with a state 
that is annihilated by half of the fermionic operators appearing in the 
supercharges (\ref{Qs}). There are only two possible states, $\ket{+}$ and 
$\ket{-}$, satisfying
\bea
\bar{\lambda}_1\ket{+}= \bar{\lambda}_4\ket{+}=\lambda_2\ket{+}=\lambda_3
\ket{+}&=&0,\nonumber\\
\lambda_1\ket{-}= \lambda_4\ket{-}=\bar{\lambda}_2\ket{-}=\bar{\lambda}_3
\ket{-}&=&0\ .
\label{choices}
\eea
It turns out that it is impossible for any other spin choices to preserve
all four supersymmetries; the wave function would have to satisfy 
incompatible differential equations.  From the quantum
mechanics superalgebra, partial supersymmetry breaking is not allowed
\cite{Witten:1982df}.

>From (\ref{Qs}), the wave function corresponding to the first state must 
satisfy
\be
(p-A)_{w^1} \phi_+=(p-A)_{\bar{w}^2}\phi_+=0\ .
\ee
Using (\ref{potw}) for the potential, the solution is
\be\label{wave1}
\phi_+=e^{-\frac{1}{2 \pi R_2 R_3}\frac{h}{g_s}\left(|w^1|^2+|w^2|^2\right)} 
F(\bar{w}^1,w^2)\ .
\ee
As in the $T^2$ example, when going around the torus, the wave function picks 
up a phase given by the gauge transformations ({\it cf}. Eqs (\ref{gaugetr}) and 
(\ref{period})). When $x^5\rightarrow x^5+2\pi R_2$, we get the following 
condition on the function $F(\bar{w}^1,w^2)$
\be\label{Fgauge}
 F(\bar{w}^1+\frac{h_1}{\sqrt{2h(h-h_2)}}\pi R_2,w^2+
\frac{h-h_2}{\sqrt{2h(h-h_2)}}\pi R_2)=e^{-\frac{v}{R_3}
\frac{h}{g_s} -\frac{\pi R_2}{2R_3}\frac{h}{g_s}}\,  F(\bar{w}^1,w^2)
\ee
where
\be
v=\frac{1}{\sqrt{2h(h-h_2)}}\left(h_1 \bar{w}^1+(h-h_2) w^2\right), 
\qquad u=\frac{1}{\sqrt{2h(h-h_2)}}\left((h-h_2)\bar{w}^1-h_1 w^2\right)
\label{uvdef}
\ee
(we defined $u$ for future use).

In a similar fashion to the $T^2$ example, we define the function $G$ as
\be
F(\bar{w}^1,w^2)=e^{\frac{1}{2\pi R_2 R_3}\frac{h}{g_s}
((\bar{w}^1)^2+(w^2)^2)} G(\bar{w}^1,w^2)
\ee
which should be periodic when $\bar{w}^1\rightarrow w^1+
\frac{h_1}{\sqrt{2h(h-h_2)}}\pi R_2$ and $w^2\rightarrow
\frac{(h-h_2}{\sqrt{2h(h-h_2)}}\pi R_2$, or, using the variables $u$ and 
$v$ in (\ref{uvdef}), $v\rightarrow v+\pi R_2$ and $u\rightarrow u$. So we 
can decompose $G$ in terms of Fourier modes. Putting  everything together, our 
wave function is
\be
\phi_+=e^{-\frac{1}{2 \pi R_2 R_3}\frac{h}{g_s}\left(|w^1|^2+|w^2|^2\right)
+\frac{1}{2\pi R_2 R_3}\frac{h}{g_s}((\bar{w}^1)^2+(w^2)^2)}
\sum_{m} e^{2im\frac{v}{R_2}} g_m(u)
\label{solphi+}
\ee
where $g_m(u)$ is to be determined from the other periodicity conditions. 

\begin{figure}[t]
\includegraphics[scale=0.7]{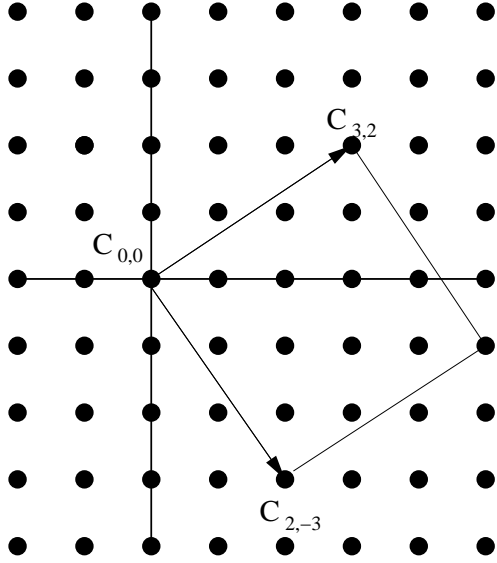}
\caption{\label{f:fund} Lattice of coefficients $C_{m,n}$. For 
$\frac{h_1}{g_s}=2$, 
$\frac{h_2}{g_s}=3$, the independent coefficients are those that lie inside 
the fundamental cell. The number of BPS states in this case is 
$13=\left(\frac{h_1}{g_s}\right)^2 + \left(\frac{h_2}{g_s}\right)^2$.}
\end{figure}

When going around the torus in the $x^8$ direction, i.e. when 
$x^8\rightarrow x^8 + 2 \pi R_2$, the condition on the wave function forces 
$g_m(u)$ to have periodicity $\pi R_2$. So our Fourier mode decomposition in 
(\ref{solphi+}) is a double sum, i.e.
\be
\phi_+=e^{-\frac{1}{2 \pi R_2 R_3}\frac{h}{g_s}\left(|w^1|^2+|w^2|^2\right)
+\frac{1}{2\pi R_2 R_3}\frac{h}{g_s}((\bar{w}^1)^2+(w^2)^2)}\sum_{m,n}C_{m,n} 
e^{2i\frac{ (mv+nu)}{R_2}} 
\label{sol2phi+}
\ee
Finally, similar to the $T^2$ case, going around the torus in the $x^6$ 
and $x^9$ gives us two recursive relations for the coefficients $C_{m,n}$, 
of the form
\bea
C_{m+\frac{h_2}{g_s}, n+\frac{h_1}{g_s}}=C_{m,n} e^{-\pi\frac{R_3}{R_2} 
\frac{h}{g_s}-2\pi \frac{R_3}{R_2}\frac{(h_2 m+h_1 n)}{h}} \nonumber\\
C_{m+\frac{h_1}{g_s}, n-\frac{h_2}{g_s}}=C_{m,n} e^{-\pi\frac{R_3}{R_2} 
\frac{h}{g_s}-2\pi \frac{R_3}{R_2}\frac{(h_1 m-h_2 n)}{h}}\ .
\label{recursion}
\eea
The series defined by these recursive relations converges for any sign of 
$h_1$ and $h_2$ and factorizes into theta functions as
\bea
\phi_+&=&\sum_{k,l}D_{k,l}
\exp\left[-\frac{1}{2 \pi R_2 R_3}\frac{h}{g_s}\left(|w^1|^2+|w^2|^2
-(\bar{w}^1)^2-(w^2)^2\right)\right] \nonumber\\
&&\times
\vartheta\left[\begin{array}{c} \frac{g_s (h_2 k+h_1 l)}{h^2}\\0\end{array}
\right] \left( \frac{h_2 v+h_1 u}{\pi R_2 g_s},\frac{ihR_3}{2 R_2 g_s}
\right) \vartheta\left[\begin{array}{c} \frac{g_s (h_1 k-h_2 l)}{h^2}\\0
\end{array}
\right] \left( \frac{h_1 v-h_2 u}{\pi R_2 g_s},\frac{ihR_3}{2 R_2 g_s}
\right) \label{theta4}\eea
after solving the recursion, where the sum on $k,l$ is over points in the
unit cell for $m,n$ as in (\ref{sol2phi+}) (see figure \ref{f:fund}).

If instead of working with the first choice of ground state in 
(\ref{choices}) we start with the second possibility, we get a similar 
solution to (\ref{sol2phi+}), but the recursive relations are such that the 
series doesn't converge for any sign of $h_1$ and $h_2$.  At first glance,
the fact that the $\ket{+}$ state is normalizable for any magnetic field,
while the $\ket{-}$ is not, appears contradictory with the results for a
particle on $T^2$.  Physically, we expect that, as in the particle case,
a change of sign of the magnetic field would be compensated by a change of
spin.  This is indeed still the case in the current scenario; our 
change of basis (\ref{lambdaferms}) reverses the physical spin of the 
string (given by the $\psi$ and $\phi$ variables) as the field is reversed.

The number of ground states is given by the number of independent 
coefficients. This number is equal to $\frac{h^2}{g_s^2}=
\frac{h_1^2+h_2^2}{g_s^2}=f_2^2+f_1^2$, as can be seen from figure 
\ref{f:fund}.

\begin{figure}[t]
\includegraphics[scale=0.7]{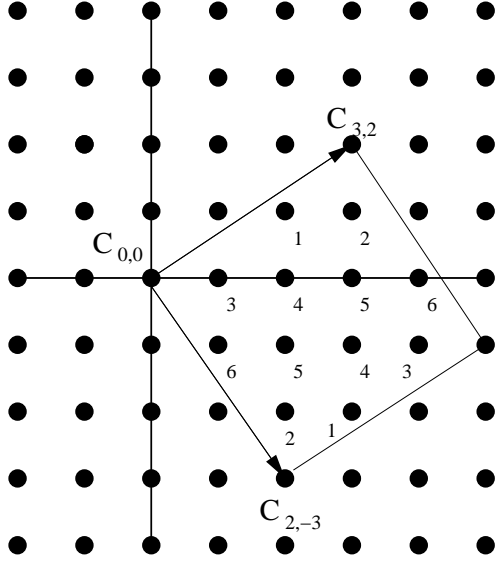}
\caption{\label{f:orb}  The lattice points inside the unit cell for
the same case as in figure \ref{f:fund}; numbered points are
identified under the orientifold reflection followed by lattice
translations (equivalent to identifying point related by a reflection
with respect to the center of the lattice). }
\end{figure}

The orientifold projection forces $C_{m,n}=C_{-m,-n}$. Then, some of the 
elements in the fundamental lattice are related to one another. For the case 
$\frac{h_1}{g_s}=2$, $\frac{h_2}{g_s}=3$, we indicate in figure \ref{f:orb} 
all those 
coefficients that are related after the orientifold projection and a 
translation along the lattice basis vectors. In this case, the number of 
independent coefficients is $7$ ($6$ interior points, plus the origin).  

There is no general formula for the number of ground states that
survive after the projection, but nevertheless there are only a finite
number of cases to consider in the string compactification. 
The 3-form flux carries D3-brane charge,
as can be seen from the equation for the warp factor
(\ref{warp}). Tadpole cancellation then imposes conditions on this
flux of the form \cite{Frey:2002hf} $g_s (f_1^2+f_2^2)\le 8$. 
Besides, the self duality condition on the 3-form flux 
requires $g_s \ge \frac{1}{min f_{1,2}}$, leaving only the possibilities
considered in table \ref{t:counting} (the number of ground states for the pair
$\left(\frac{h_1}{g_s}, \frac{h_2}{g_s}\right)=(1,0)$, for example, is
equal to that for $\left(\frac{h_1}{g_s},
\frac{h_2}{g_s}\right)=(0,1)$).

\begin{table}[t]
\begin{center}
\begin{tabular}{|c|c|c|}
\hline
$\left(\frac{h_1}{g_s}, \frac{h_2}{g_s}\right)$ & \# indep. $C_{m,n}$  
&  \# indep. $C_{m,n}$  \\
 & before orientifold & after orientifold\\
\hline
$(0,1)$ & 1 & 1 \\	
$(0,2)$ & 4 & 4 \\
$(0,3)$ & 9 & 5 \\
$(0,4)$ & 16 & 10 \\
$(1,1)$ & 2 & 2 \\
$(1,2)$ & 5 & 3 \\
$(2,3)$ & 13 & 7 \\
$(3,3)$ & 18 & 10 \\
$(4,4)$ & 32 & 17 \\
\hline
\end{tabular}
\end{center}
\caption{\label{t:counting} 
Possible combinations of 3-form fluxes, and the number of ground states 
obtained before and after the orientifold projection.}
\end{table}

There is one final issue to consider; in some cases, a point $(m,n)$
maps to itself under the $\Z$ projection followed by several recursion
relations among the Fourier coefficients.  In order to verify that those 
points truly survive the orientifold, we should check that the 
necessary recursion relations truly give $C_{-m,-n}=C_{m,n}$.  First, we note
that, in this case, $\pm (m,n)$ are separated by integer numbers of the
recursion shift vectors:
\be\label{orbshifts}
(m,n) -q(h_2/g_s,h_1/g_s)-p(h_1/g_s,-h_2/g_s) = (-m,-n)\ ,\quad
p,q\in \mathbf{Z}\ .\ee
Since the recursion relations (\ref{recursion}) are independent of 
position along the orthogonal shift vector, we have then
\bea
C_{-m,-n} &=&\exp\left\{-\frac{\pi R_3}{R_2} \left[(p+q)\frac{h}{g_s}
-\frac{2h_2}{h} \sum_{i=0}^{q-1} 
\left( m-i\frac{h_2}{g_s}\right)
-\frac{2h_1}{h} \sum_{i=0}^{q-1} 
\left( n-i\frac{h_1}{g_s}\right)\right.\right.\nonumber\\
&&\left.\left.-\frac{2h_1}{h} \sum_{j=0}^{p-1} 
\left( m-j\frac{h_1}{g_s}\right)
+\frac{2h_2}{h} \sum_{j=0}^{p-1} 
\left( n+j\frac{h_2}{g_s}\right)\right]\right\} C_{m,n}\ .\label{check}
\eea
Carrying out the arithmetic sums and using (\ref{orbshifts}) shows
easily that the exponent indeed vanishes.

\section{Other BPS States}\label{s:extension}

There are other BPS states that we can consider.  In this section, we 
will find BPS states localized at singularities of the D-string moduli
space, show how to count the states of F-strings, and discuss briefly
the problem of threshhold bound states.\footnote{We are grateful to 
our referee for comments that led to the analysis in sections
\ref{ss:singularity} and \ref{ss:bound}.}

\subsection{Singularities in Moduli Space}\label{ss:singularity}
As the D-string moves on the torus, it encounters two types of singularities.
First, at $x^5=\cdots =x^9=0$, it passes through two orientifold planes
and doubles back on itself.  (Note that the D-string does not end on an 
O3-plane; this is because, with vanishing flux, it is T-dual to a type I
D5-brane, which always has 2 CP indices.)
Also, if tadpole cancellation so requires, the
string could intersect a D3-brane (which we specifically ignored in our
calculation).  These are direct analogs of the singularities in the
moduli space of type I and heterotic 5-branes mentioned in 
\cite{Witten:1996gx}.

To count states localized at the singularities, we should be able to 
ignore the 3-form
background.  The reason is the localized states do not have zero modes that
can move through the fluxes.  Therefore, the counting should be the same as
without 3-forms, which is dual to the $SO(32)$ heterotic 
5-brane.\footnote{For an $E_8$ application, see \cite{Keurentjes:2002dc}.}  
Translating the results of \cite{Porrati:1996vj,Sethi:1997kj} to IIB 
language, we find that there are no localized states at the O3-planes and 
that there is one BPS multiplet attached to each D3-brane.

\subsection{Counting for an F-string}\label{ss:duality}

We can obtain the number of BPS states for an F-string by starting with
the Green-Schwarz action for superstrings and accounting for the background
fields; because of supersymmetry, we expect it to suffice for the 
calculation of our index.  However, it is easier to proceed 
by applying S-duality and rotating the background fields,
and then using the results from the previous section. We want to find
the number of BPS states for an F-string wrapped in the 7-direction,
so we will S-dualize the background, and then rotate in the 4-7 plane.

Under an $SL(2,{\bf Z})$ transformation, 
\be\label{Sdual} \tau \rightarrow
\frac{a\tau+b}{c\tau+d}, \quad \quad \quad  \pmatrix{H_{(3)} \cr
F\,_{(3)}\cr} \rightarrow \pmatrix{d & c \cr b & a\cr}
\pmatrix{H_{(3)} \cr F_{(3)}\cr}\ .\ee 
An S-duality is $\tau
\rightarrow -\frac{1}{\tau}$. Then,  $H_{(3)} \rightarrow F_{(3)}$ and
$F_{(3)} \rightarrow -H_{(3)}$.  So $h_1$ and $h_2$ from the previous
sections get mapped to $f_1$ and  $f_2$ respectively. Using the duality
condition on the 3-form flux (\ref{selfduality}), the action of
S-duality is $h_1 \rightarrow -\frac{h_2}{g_s}$ and  $h_2 \rightarrow
\frac{h_1}{g_s}$.

A rotation in the 4-7 plane interchanges $h_1$ with $h_2$ again. From
(\ref{fluxes}), we see that a rotation of $\pi /2$ gives $h_1 \rightarrow
-h_2$ and $h_2 \rightarrow h_1$. A rotation of $- \pi /2$ does the
same thing with opposite signs.

Then, the combined action of an S-duality and a rotation, gives
$h_{1,2} \rightarrow -\frac{h_{1,2}}{g_s}$ in one case and $h_{1,2}
\rightarrow \frac{h_{1,2}}{g_s}$ for the opposite rotation. Then, the
number of BPS states  of the rotated F-string is just
$h_1^2+h_2^2$ plus one state at each D3-brane. 
(In the D-string, the number of states is
$f_1^2+f_2^2$ plus one for each D3-brane.)  So the total number of BPS
multiplets for a given BPS charge should be $f_1^2+f_2^2+h_1^2+h_2^2+
2N_{D3}.$

\subsection{Bound States}\label{ss:bound}

So far we have considered only states of  one D- or 
F-string at a time, which are the states of minimal BPS electric charge.
However, it would be very interesting to consider the BPS spectrum of 
multiple strings, since there could be threshhold bound states.  We will
largely leave this question for the future, but we can make some comments.

Consider, for example, the case of two BPS charges.  If they are widely
separated in the noncompact dimensions, then the spectrum should just be
the direct product of the spectra of the individual charges (appropriately
symmetrized).  As the two charges become coincident, we expect that there
would, in addition, be BPS multiplets associated with threshhold bound states.
The counting of multiplets breaks down according to the nature of the 
strings.  A bound state of two D-strings, for example, would have twice the
charge but would be otherwise identical to our previous analysis, so there
would be four times as many nonlocalized states.  A bound state of a
D-string and F-string seems more complicated in that the two strings feel
different fluxes.

Another question that we leave to future work is the nature of the bound
states.  For two D-strings, we really should take into account the 
non-Abelian nature of the worldvolume theory; we can do this using the
D-brane action of \cite{Myers:1999ps}, and supersymmetrizing.  It is possible that the bound
state is a ``polarized'' configuration, or it is possible that there are
polarized and unpolarized bound states.  We should also mention that
we could start with, instead of the action of \cite{Myers:1999ps}, 
super-Yang-Mills with a superpotential due to the
3-forms along with the appropriate velocity coupling to the vector potential. 
The problem is to guess how the superpotential depends on the 3-form flux.  

\section{Conclusions}\label{s:conclude}
We were able to count the number of short multiplets for both a D- and
an F-string wrapped on a compact direction of a $T^6/{\bf Z}_2$
orientifold,  on a background containing constant 3-form fluxes.  We
showed that the number of states is proportional to the square of the
units of 3-form flux enclosed on $T^3$ cycles inside the $T^6$. Modulo
algebraic complications, the counting of ground states follows two
copies of that of a superparticle on a torus with constant
perpendicular magnetic field.

This is a nice result obtained from the $\kappa$-symmetric action for
a D1-brane in background fluxes, which was our starting point. We
showed that the Hamiltonian obtained from this action preserved the
amount of supersymmetries expected, and from the supercharges we
obtained the supersymmetric ground states. By S-duality, we were able
to find the corresponding number of states for the F-string case.

One of the motivations of this work was the mismatch found in
\cite{Frey:2002hf} between the number of states carrying minimal
electric (D- and F-strings) versus magnetic (bound state of D5- and
NS5-brane) charges in the $\N=3$ theory. It was then conjectured that
the sum of the BPS states of the D- and F-string should be equal to
those of the bound state of 5-branes. In this paper we computed the
former, while the latter is left for future work.

It is our hope that we have introduced a useful piece of technology for
the study of string and D-brane physics in flux backgrounds.
   
\begin{acknowledgments}
We are greatly indebted to J. Polchinski for guidance throughout this
project. We would like to thank also E. D'Hoker for useful discussions and
our referee for very helpful comments. 
This work was supported by NSF grant PHY97-22022.
\end{acknowledgments}
 
\appendix
\section{Conventions}\label{a:conventions}

In the following, Greek
indices refer to $0,1,2,3$ directions, while Roman indices $m,n,..$
refer to $4$ to $9$ directions.  

There are two
spacetime Majorana-Weyl spinors $\Theta^1$ and $\Theta^2$ in the 
$\kappa$-symmetric action. The
$\kappa$-symmetry can be
used to fix a gauge, keeping only 16 of those 32 degrees of
freedom. We used the gauge $\Theta^2=0$.
 
The spinor $\Theta$ in Eqs (\ref{action1}) and (\ref{expansionIIB})
below, is a Majorana-Weyl spinor in a Dirac basis: \be\label{thetaferm}
\Theta=\pmatrix{0 \cr \theta \cr}.  \ee It can be decomposed into a
Majorana-Weyl spinor in $SO(1,1)$ and a Majorana-Weyl spinor in
$SO(8)$ as in Eq.(\ref{decomp}).   A basis of Majorana-Weyl spinors in
$SO(8)$, given by spins in the complex planes defined in
eq. (\ref{zcomplex})  is     \bea  u_1=\ket{++++}+\ket{----} 
&,&v_1=\ket{+-++}-\ket{-+--} \nonumber\\
u_2=i\left(\ket{++++}-\ket{----}\right) &,
&v_2=i\left(\ket{+-++} +\ket{-+--}\right) \nonumber\\
u_3=\ket{++--}-\ket{--++} &,&v_3=\ket{-+++}+\ket{+---}
\nonumber\\ 
u_4=i\left(\ket{++--}+\ket{--++}\right) &,
&v_4=i\left(\ket{-+++} -\ket{+---}\right) \nonumber\\
u_5=\ket{+-+-}+\ket{-+-+} &,&v_5=\ket{+++-}-\ket{---+}
\nonumber\\ 
u_6=i\left(\ket{+-+-}-\ket{-+-+}\right)
&,&v_7=i\left(\ket{++-+} +\ket{--+-}\right) \nonumber\\
u_7=\ket{+--+}-\ket{-++-} &, &v_7=\ket{++-+}-\ket{--+-}
\nonumber\\ 
u_8=i\left(\ket{+--+}-\ket{-++-}\right)
&,&v_8=i\left(\ket{++-+} -\ket{--+-}\right) \ .
\label{so8basis}\eea
The $u_\alpha$ ($v_{\da}$) form an $\mathbf{8}$ ($\mathbf{8}^\prime$).

The expansions of the spacetime fields in terms of the world-volume
spinor $\Theta$ in the gauge $\Theta^2=0$ are, for constant
dilaton-axion \cite{Grana:2002tu}
\begin{eqnarray}
\bm{e}_m^a&=&e_m^a + \frac{i}{8} \overline{\Theta} \Gamma^{abc}
\Theta w_{mbc}-  \frac{i}{16}  \overline{\Theta} \Gamma^{anp} \Theta
H_{mnp} \nonumber\\
\bm{e}_{\mu }^a&=& \frac{i}{2}
(\overline{\Theta}\Gamma^a)_{\mu} \nonumber \\    {\bf e}_m^{\mu}&=& 0
\nonumber \\ 
{\bf e}_{\mu}^{\alpha}&=& \delta_{\mu}^{\alpha} \nonumber
\\  
\bm{\Phi}&=& \phi- \frac{i}{48}  \overline{\Theta}
\Gamma^{pqr} \Theta H_{pqr} \nonumber\\ 
\bm{B}_{mn} &=& B_{mn}
+\frac{i}{4}\overline{\Theta} \Gamma^{ab}\,_{[m} \Theta \,w_{n]ab} -
\frac{i}{8}\overline{\Theta} \Gamma^{pq}\,_{[m} \Theta \,H_{n]pq}
\nonumber\\ 
\bm{B}_{m\mu}&=& \frac{i}{2} (\overline{\Theta}
\Gamma_m)_{\mu} \nonumber\\ 
\bm{C}_{mn}&=& C_{mn}-
\frac{i}{8}\overline{\Theta} \Gamma^{pq}\,_{[m} \Theta \,F'_{n]pq}+ C
B_{mn}|_{\Theta^2} \nonumber\\ 
\bm{C}_{m\mu}&=& C B_{m\mu}
\nonumber\\
\bm{C}&=& C -\frac{i}{48}\overline{\Theta}
\Gamma^{pqr} \Theta F'_{pqr}\ .
\label{expansionIIB} 
\end{eqnarray}
We have introduced a factor of $i$ in each fermion bilinear, so that the
action (\ref{action1}) 
matches usual quantum field theory conventions and gives a real 
anticommutator.

\section{Corrections to the quantum mechanics}\label{a:assumptions}
Here we will estimate a number of possible corrections to the Hamiltonian
given in section \ref{ss:action} and argue that they will not change
the counting of BPS states.  Because a single long multiplet could only be
formed from four short multiplets (of differing spins), we anticipate
that it would be difficult to lift the supersymmetric states given in
section \ref{s:wavefunction}; this is essentially the argument of
\cite{Witten:1982df}.
We will mainly here be concerned if any
of the long multiplets could be made BPS by corrections to our 
Hamiltonian, eqn (\ref{hamil}).  We work in the framework of 
perturbation theory and concentrate on the bosonic terms.  We take all
the coordinate radii to be similar $\sim R$.

First we note the structure of the Hilbert space once we include excited 
states.  The center of mass modes considered in the text have excited
states with energy of the order of the magnetic field, $\sim \ap/R^3$.  Then
there are modes that move around the D-string, which form a tower of 
supersymmetric
harmonic oscillators with frequency $\sim 1/R$.  Each oscillator has
a unique supersymmetric ground state, and we don't expect perturbations to
the Hamiltonian to break the supersymmetry.\footnote{We should note that
the oscillator amplitudes are not periodic, unlike center of mass modes.
This is clear because these modes represent bending of the string, not
position.}

Just like the center of mass modes, the oscillators couple to the 3-form;
we treat this as a perturbation.  Because the perturbation couples different
oscillators, it appears only in second order perturbation theory.  Since
the magnetic field is of order $\ap/R^3$, the energy shift is of order 
$\ap{}^2/R^5$.  The 3-form flux $F_{mn7}$ also couples to the oscillators; this
perturbs the excited state energies by $\sim \ap{}^3/R^7$.  There are also
higher derivative terms in the D-brane action, coming from the
Born-Infeld determinant.  One typical term is $(\dot X \del_4 X)^2$.  This
can contribute at first order in perturbation theory, giving an energy
shift $\ap/R^2$ times the energy of the state.  

We also briefly consider bulk supergravity effects.  At large but finite 
radius, 
there is a metric warp factor; it's lowest order contribution is 
$\ap{}^2/R^4$ times the energy of the state.  The interaction with 
bulk modes is a little trickier to understand.  The massless moduli of
the compactification should not affect the BPS spectrum, since the same
supercharges are preserved in the spacetime.  The BPS particle mass and 
excited state energies will change, however.  The massive scalars of
the 4D effective theory, such as the dilaton, should also not change the
BPS spectrum.  One argument for this is that the dilaton enters into the
Hamiltonian only through the mass of the BPS particle (when one takes 
into account the normalization of the fermions), which affects only the 
excited states of the center of mass modes.  Another argument is that there
are some cancellations among different terms in the perturbation theory
for the dilaton.

Finally, we should note that the 4D gauge fields \emph{can} change the 
BPS spectrum rather violently.  For example, a BPS particle in the field
of an oppositely charged particle should have no supersymmetric ground
states.

\bibliographystyle{utcaps2} \bibliography{susyqm}

\providecommand{\href}[2]{\texttt{#2}}\begingroup\raggedright\begin{thebibliog%
raphy}{10}

\bibitem{Randall:1999ee}
L.~Randall and R.~Sundrum, ``A large mass hierarchy from a small extra
  dimension,'' {\em Phys. Rev. Lett.} {\bf 83} (1999) 3370--3373,
\href{http://www.arXiv.org/abs/hep-ph/9905221}{ hep-ph/9905221}.

\bibitem{Randall:1999vf}
L.~Randall and R.~Sundrum, ``An alternative to compactification,'' {\em Phys.
  Rev. Lett.} {\bf 83} (1999) 4690--4693,
\href{http://www.arXiv.org/abs/hep-th/9906064}{ hep-th/9906064}.

\bibitem{Becker:1996gj}
K.~Becker and M.~Becker, ``M-Theory on Eight-Manifolds,'' {\em Nucl. Phys.}
  {\bf B477} (1996) 155--167,
\href{http://www.arXiv.org/abs/hep-th/9605053}{ hep-th/9605053}.

\bibitem{Mayr:2000zd}
P.~Mayr, ``Stringy world branes and exponential hierarchies,'' {\em JHEP} {\bf
  11} (2000) 013,
\href{http://arXiv.org/abs/hep-th/0006204}{ hep-th/0006204}.

\bibitem{Gukov:1999ya}
S.~Gukov, C.~Vafa, and E.~Witten, ``CFT's from Calabi-Yau four-folds,'' {\em
  Nucl. Phys.} {\bf B584} (2000) 69--108,
\href{http://www.arXiv.org/abs/hep-th/9906070}{ hep-th/9906070}.

\bibitem{Dasgupta:1999ss}
K.~Dasgupta, G.~Rajesh, and S.~Sethi, ``M theory, orientifolds and G-flux,''
  {\em JHEP} {\bf 08} (1999) 023,
\href{http://www.arXiv.org/abs/hep-th/9908088}{ hep-th/9908088}.

\bibitem{Greene:2000gh}
B.~R. Greene, K.~Schalm, and G.~Shiu, ``Warped compactifications in M and F
  theory,'' {\em Nucl. Phys.} {\bf B584} (2000) 480--508,
\href{http://www.arXiv.org/abs/hep-th/0004103}{ hep-th/0004103}.

\bibitem{Grana:2000jj}
M.~Grana and J.~Polchinski, ``Supersymmetric three-form flux perturbations on
  AdS(5),'' {\em Phys. Rev.} {\bf D63} (2001) 026001,
\href{http://www.arXiv.org/abs/hep-th/0009211}{ hep-th/0009211}.

\bibitem{Gubser:2000vg}
S.~S. Gubser, ``Supersymmetry and F-theory realization of the deformed conifold
  with three-form flux,''
\href{http://www.arXiv.org/abs/hep-th/0010010}{ hep-th/0010010}.

\bibitem{Giddings:2001yu}
S.~B. Giddings, S.~Kachru, and J.~Polchinski, ``Hierarchies from fluxes in
  string compactifications,''
\href{http://www.arXiv.org/abs/hep-th/0105097}{ hep-th/0105097}.

\bibitem{Becker:2002nn}
K.~Becker, M.~Becker, M.~Haack, and J.~Louis, ``Supersymmetry breaking and
  alpha' corrections to flux induced potentials,''
\href{http://arXiv.org/abs/hep-th/0204254}{ hep-th/0204254}.

\bibitem{Verlinde:1999fy}
H.~Verlinde, ``Holography and compactification,'' {\em Nucl. Phys.} {\bf B580}
  (2000) 264--274,
\href{http://www.arXiv.org/abs/hep-th/9906182}{ hep-th/9906182}.

\bibitem{Kachru:2002he}
S.~Kachru, M.~Schulz, and S.~Trivedi, ``Moduli stabilization from fluxes in a
  simple IIB orientifold,''
\href{http://arXiv.org/abs/hep-th/0201028}{ hep-th/0201028}.

\bibitem{Frey:2002hf}
A.~R. Frey and J.~Polchinski, ``N = 3 warped compactifications,''
\href{http://arXiv.org/abs/hep-th/0201029}{ hep-th/0201029}.

\bibitem{Andrianopoli:2002rm}
L.~Andrianopoli, R.~D'Auria, S.~Ferrara, and M.~A. Lledo, ``Super Higgs effect
  in extended supergravity,''
\href{http://arXiv.org/abs/hep-th/0202116}{ hep-th/0202116}.

\bibitem{Andrianopoli:2002mf}
L.~Andrianopoli, R.~D'Auria, S.~Ferrara, and M.~A. Lledo, ``Gauging of flat
  groups in four dimensional supergravity,''
\href{http://arXiv.org/abs/hep-th/0203206}{ hep-th/0203206}.

\bibitem{Andrianopoli:2002aq}
L.~Andrianopoli, R.~D'Auria, S.~Ferrara, and M.~A. Lledo, ``Duality and
  spontaneously broken supergravity in flat backgrounds,''
\href{http://arXiv.org/abs/hep-th/0204145}{ hep-th/0204145}.

\bibitem{D'Auria:2002tc}
R.~D'Auria, S.~Ferrara, and S.~Vaula, ``N = 4 gauged supergravity and a IIB
  orientifold with fluxes,''
\href{http://arXiv.org/abs/hep-th/0206241}{ hep-th/0206241}.

\bibitem{Ferrara:2002bt}
S.~Ferrara and M.~Porrati, ``N = 1 no-scale supergravity from IIB
  orientifolds,''
\href{http://arXiv.org/abs/hep-th/0207135}{ hep-th/0207135}.

\bibitem{Cederwall:1997ri}
M.~Cederwall, A.~von Gussich, B.~E.~W. Nilsson, P.~Sundell, and A.~Westerberg,
  ``The Dirichlet super-p-branes in ten-dimensional type IIA and IIB
  supergravity,'' {\em Nucl. Phys.} {\bf B490} (1997) 179--201,
\href{http://arXiv.org/abs/hep-th/9611159}{ hep-th/9611159}.

\bibitem{Bergshoeff:1997tu}
E.~Bergshoeff and P.~K. Townsend, ``Super D-branes,'' {\em Nucl. Phys.} {\bf
  B490} (1997) 145--162,
\href{http://arXiv.org/abs/hep-th/9611173}{ hep-th/9611173}.

\bibitem{Fradkin:1991nr}
E.~H. Fradkin, ``Field theories of condensed matter systems,''. Redwood City,
  USA: Addison-Wesley (1991) 350 p. (Frontiers in physics, 82).

\bibitem{Witten:1982df}
E.~Witten, ``Constraints on supersymmetry breaking,'' {\em Nucl. Phys.} {\bf
  B202} (1982)
253.

\bibitem{Casalbuoni:1976tz}
R.~Casalbuoni, ``The classical mechanics for Bose-Fermi systems,'' {\em Nuovo
  Cim.} {\bf A33} (1976)
389.

\bibitem{Polchinski:1998rq}
J.~Polchinski, ``String theory. Vol. 1: An introduction to the bosonic
  string,''. Cambridge, UK: Univ. Pr. (1998) 402 p.

\bibitem{deWit:1998tk}
B.~de~Wit, K.~Peeters, and J.~Plefka, ``Superspace geometry for supermembrane
  backgrounds,'' {\em Nucl. Phys.} {\bf B532} (1998) 99--123,
\href{http://arXiv.org/abs/hep-th/9803209}{ hep-th/9803209}.

\bibitem{Millar:2000ib}
K.~Millar, W.~Taylor, and M.~Van~Raamsdonk, ``D-particle polarizations with
  multipole moments of higher- dimensional branes,''
\href{http://arXiv.org/abs/hep-th/0007157}{ hep-th/0007157}.

\bibitem{Grana:2002tu}
M.~Gra\~na, ``D3-brane action in a supergravity background: The fermionic
  story,''
\href{http://arXiv.org/abs/hep-th/0202118}{ hep-th/0202118}.

\bibitem{Polchinski:1998rr}
J.~Polchinski, ``String theory. Vol. 2: Superstring theory and beyond,''.
  Cambridge, UK: Univ. Pr. (1998) 531 p.

\bibitem{Schwarz:1983wa}
J.~H. Schwarz and P.~C. West, ``Symmetries and transformations of chiral N=2 D
  = 10 supergravity,'' {\em Phys. Lett.} {\bf B126} (1983)
301.

\bibitem{Schwarz:1983qr}
J.~H. Schwarz, ``Covariant field equations of chiral N=2 D = 10 supergravity,''
  {\em Nucl. Phys.} {\bf B226} (1983)
269.

\bibitem{Bergshoeff:1997kr}
E.~Bergshoeff, R.~Kallosh, T.~Ortin, and G.~Papadopoulos, ``Kappa-symmetry,
  supersymmetry and intersecting branes,'' {\em Nucl. Phys.} {\bf B502} (1997)
  149--169,
\href{http://arXiv.org/abs/hep-th/9705040}{ hep-th/9705040}.

\bibitem{Kallosh:1998ky}
R.~Kallosh, ``Worldvolume supersymmetry,'' {\em Phys. Rev.} {\bf D57} (1998)
  3214--3218,
\href{http://arXiv.org/abs/hep-th/9709069}{ hep-th/9709069}.

\bibitem{Witten:1996gx}
E.~Witten, ``Small Instantons in String Theory,'' {\em Nucl. Phys.} {\bf B460}
  (1996) 541--559,
\href{http://arXiv.org/abs/hep-th/9511030}{ hep-th/9511030}.

\bibitem{Keurentjes:2002dc}
A.~Keurentjes and S.~Sethi, ``Twisting E8 five-branes,'' {\em Phys. Rev.} {\bf
  D66} (2002) 046001,
\href{http://arXiv.org/abs/hep-th/0205162}{ hep-th/0205162}.

\bibitem{Porrati:1996vj}
M.~Porrati, ``How to find H-monopoles in brane dynamics,'' {\em Phys. Lett.}
  {\bf B387} (1996) 492--496,
\href{http://arXiv.org/abs/hep-th/9607082}{ hep-th/9607082}.

\bibitem{Sethi:1997kj}
S.~Sethi and M.~Stern, ``A comment on the spectrum of H-monopoles,'' {\em Phys.
  Lett.} {\bf B398} (1997) 47--51,
\href{http://arXiv.org/abs/hep-th/9607145}{ hep-th/9607145}.

\bibitem{Myers:1999ps}
R.~C. Myers, ``Dielectric-branes,'' {\em JHEP} {\bf 12} (1999) 022,
\href{http://arXiv.org/abs/hep-th/9910053}{ hep-th/9910053}.

\end{thebibliography}\endgroup

\end{document}